\documentclass[10pt]{sig-alternate}
\usepackage[font={small}]{caption, subfig}
\usepackage{fix-cm}
\usepackage{listings}
\usepackage{textcomp} 
\usepackage{layouts} 
\usepackage[nolist]{acronym}
\usepackage{graphicx}
  \DeclareGraphicsExtensions{.pdf}

\renewcommand{\footnotesize}{\scriptsize}

\newcommand{\epsep}{\rceil}
\newcommand{\olsep}{\|}
\newcommand{\nolsep}{|}
\newcommand{\ecmspace}{\,}

\newcommand{\ecm}[6]{\mbox{$\left\{{#1}\ecmspace\olsep\ecmspace {#2}\ecmspace\nolsep\ecmspace {#3}\ecmspace\nolsep\ecmspace {#4}\ecmspace\nolsep\ecmspace {#5}\right\}\ecmspace{#6}$}}
\newcommand{\ecmshort}[5]{\mbox{$\left\{{#1}\ecmspace\olsep\ecmspace {#2}\ecmspace\nolsep\ecmspace {#3}\ecmspace\nolsep\ecmspace {#4}\ecmspace\nolsep\ecmspace {#5}\right\}$}}
\newcommand{\ecmp}[5]{\mbox{$\left\{{#1}\ecmspace\epsep\ecmspace {#2}\ecmspace\epsep\ecmspace {#3}\ecmspace\epsep\ecmspace {#4}\right\}\ecmspace{#5}$}}
\newcommand{\ecmpshort}[4]{\mbox{$\left\{{#1}\ecmspace\epsep\ecmspace {#2}\ecmspace\epsep\ecmspace {#3}\ecmspace\epsep\ecmspace {#4}\right\}$}}

\begin{document}

\begin{acronym}[DVFS]
    \acro{AGU}{address generation unit}
    \acro{AVX}{advanced vector extensions}
    \acro{CL}{cache line}
    \acro{COD}{cluster-on-die}
    \acro{DCT}{dynamic concurrency throttling}
    \acro{DP}{double precision}
    \acro{DVFS}{dynamic voltage and frequency scaling}
    \acro{ECM}{execution-cache-memory}
    \acro{FMA}{fused multiply-add}
    \acro{IMCI}{initial many core instructions}
    \acro{LFB}{line fill buffer}
    \acro{LLC}{last-level cache}
    \acro{MC}{memory controler}
    \acro{NT}{non-temporal}
    \acro{NUMA}{non-uniform memory access}
    \acro{RAPL}{running average power limit}
    \acro{SIMD}{single instruction multiple data}
    \acro{SP}{single precision}
    \acro{UFS}{Uncore frequency scaling}
\end{acronym}


\setcopyright{acmcopyright}
\conferenceinfo{E2SC '16}{November 14\textsuperscript{th}, 2016, Salt Lake City, UT, USA}
\CopyrightYear{2016} 
\crdata{0-12345-67-8/90/01}  
\doi{10.475/123_4}
\isbn{123-4567-24-567/08/06}
\acmPrice{\$15.00}

\title{An ECM-based energy-efficiency optimization approach for bandwidth-limited streaming kernels on recent Intel Xeon processors}

\numberofauthors{2} 
\author{
\alignauthor
Johannes Hofmann\\
       \affaddr{Department of Computer Science}\\
       \affaddr{University of Erlangen-Nuremberg}\\
       \affaddr{Erlangen, Germany}\\
       \email{johannes.hofmann@fau.de}
\alignauthor
Dietmar Fey\\
       \affaddr{Department of Computer Science}\\
       \affaddr{University of Erlangen-Nuremberg}\\
       \affaddr{Erlangen, Germany}\\
       \email{dietmar.fey@fau.de}
}

\maketitle
\begin{abstract}
We investigate an approach that uses low-level analysis and the
execution-cache-memory (ECM) performance model in combination with tuning of
hardware parameters to lower energy requirements of memory-bound applications.
The ECM model is extended appropriately to deal with software optimizations
such as non-temporal stores.  Using incremental steps and the ECM model, we
analytically quantify the impact of various single-core optimizations and
pinpoint microarchitectural improvements that are relevant to energy
consumption.  Using a 2D Jacobi solver as example that can serve as a blueprint
for other memory-bound applications, we evaluate our approach on the four most
recent Intel Xeon E5 processors (Sandy Bridge-EP, Ivy Bridge-EP, Haswell-EP,
and Broadwell-EP).  We find that chip energy consumption can be reduced in the
range of 2.0--2.4$\times$ on the examined processors. 
\end{abstract}


%


\section{Introduction and Rel. Work}
\label{sec:intro}

For new HPC installations contribution of power usage to total system cost has
been increasing steadily over the past years 
\cite{koomey2011growth} and studies project this trend to continue \cite{fraunhofer}.
As a consequence energy-aware metrics have recently been gaining popularity.
Energy-to-solution, i.e. the amount of energy
consumed by a system to solve a given problem, is the most obvious of
these metrics and will be used as quality measure throughout this study.

Previous works view the application code as constant and instead focus their
energy optimization attempts to parameter tuning on either the runtime
environment, the hardware, or both \cite{journals/tpds/LiSSNC13}. The runtime
environment approach works by adjusting the number of active threads across
different parallel regions based on the regions' computational requirements
\cite{4407685}. Hardware parameter tuning involves trying to identify slack,
e.g. during MPI communication, and using \ac{DVFS} to lower energy consumption
in such sections of low computational intensity
\cite{Miyoshi:2002:CPS:514191.514200}; similar strategies can be applied to
OpenMP barriers \cite{4725146}.  The parameters employed by these algorithms
can be based on measurements made at runtime
\cite{Rountree:2009:AMD:1542275.1542340} or set statically \cite{5348808}. One
fact often ignored by \ac{DVFS} control software, however, is that hardware
delays caused by changing frequency states can often be significant
\cite{Mazouz2013} which can lead to diminishing returns in the real world.
Another form of hardware parameter optimization involves the tuning of hardware
prefetchers \cite{Wu:2011:CDM:2015039.2015516}.

In contrast to previous work, our approach focuses on increasing single-core
performance, primarily though software optimization.  Using the \ac{ECM}
performance model to guide optimizations such as SIMD vectorization, cache
blocking, non-temporal stores, and the use of \ac{COD} mode the bandwidth
consumption of a single core is maximized. This allows the chip to saturate
main memory bandwidth with the least number of active cores in the multi-core
scenario, thus reducing power as well as the energy consumed by the chip.  In a
second step, different hardware parameters such as the number of active cores
and their frequencies should be evaluated to further reduce energy consumption.
The approach is demonstrated on a 2D Jacobi solver, which acts as a proxy for
many memory-bound applications.  To produce significant results, all
experiments were performed on the four most recent server microarchitectures by
Intel, which make up about 91\% of HPC installations of the June 2016 Top 500
list.

The paper is organized as follows. Section~\ref{sec:blueprint} describes the
blueprint of and the reasoning behind our energy-efficiency optimization
approach.  Section~\ref{sec:ecm} introduces the \ac{ECM} performance model that
we use to guide optimizations.  Section~\ref{sec:testbed} contains an overview
of the systems used for benchmarking.  Core-level improvement efforts are
documented in Section~\ref{sec:scopt}, followed by a validation of single-core
results in Section~\ref{sec:res:sc}. Implications for the multi-core scenario are
discussed in Section~\ref{sec:res:mc}, followed by the conclusion in Section~\ref{sec:conc}.

\section{Optimization Approach}
\label{sec:blueprint}

Per definition, the bottleneck for bandwidth-bound applications is the
sustained bandwidth $b_\mathrm{s}$ imposed by the memory subsystem.  It is a
well-established fact that a single core of modern multi- and many-core systems
cannot saturate main memory bandwidth \cite{Treibig:2009}, necessitating the
use of multiple cores to reach peak performance. The number of cores
$n_\mathrm{s}$ required to sustain main memory bandwidth depends on single-core
performance: The faster a single-core implementation, the more bandwidth is
consumed by a single core. The more bandwidth is consumed by a single core, the
fewer cores are required to saturate main memory bandwidth.

\begin{figure}[!tb]
    \centering
    \includegraphics[width=0.47\linewidth]{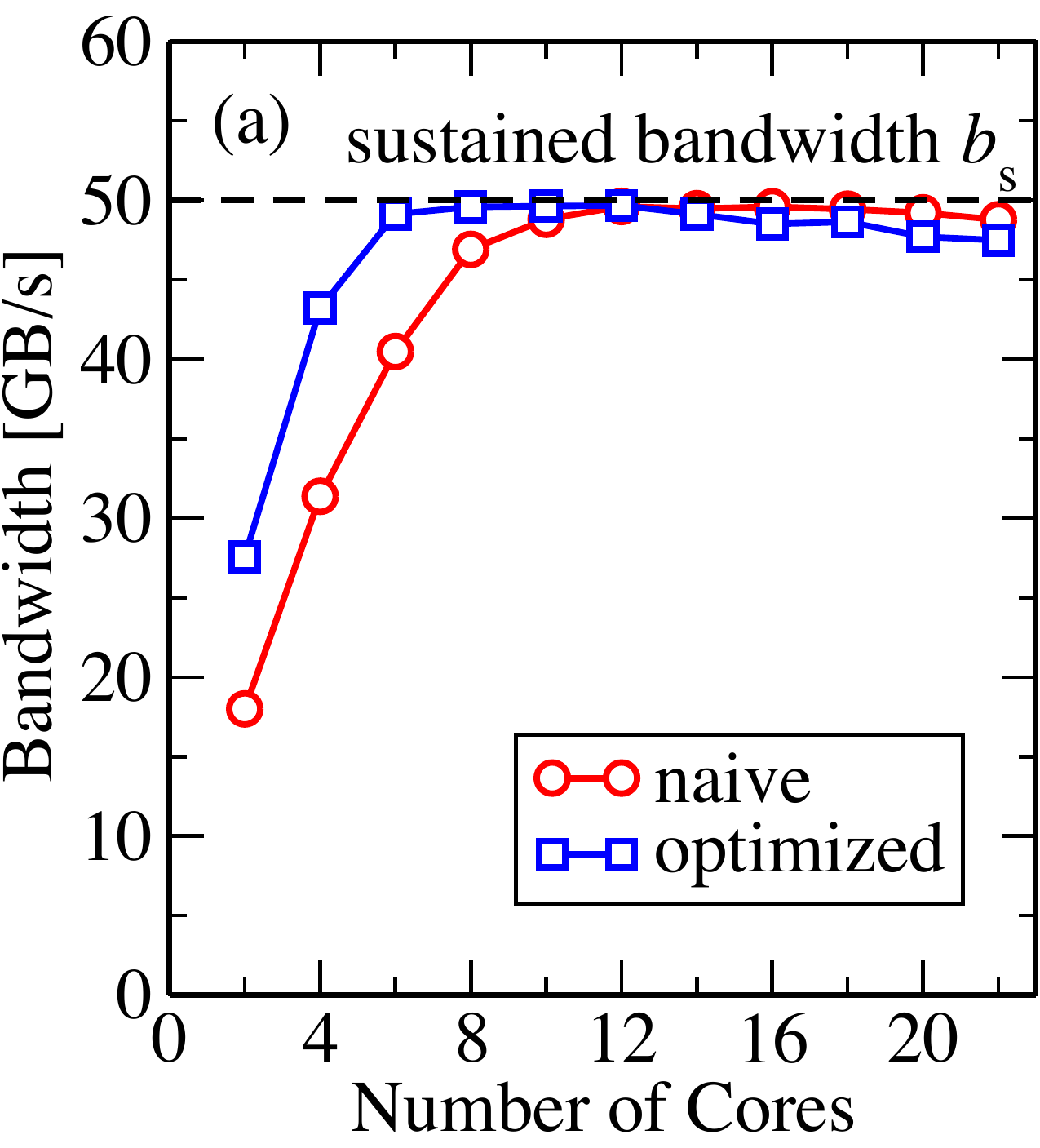}\hfill
    \includegraphics[width=0.47\linewidth]{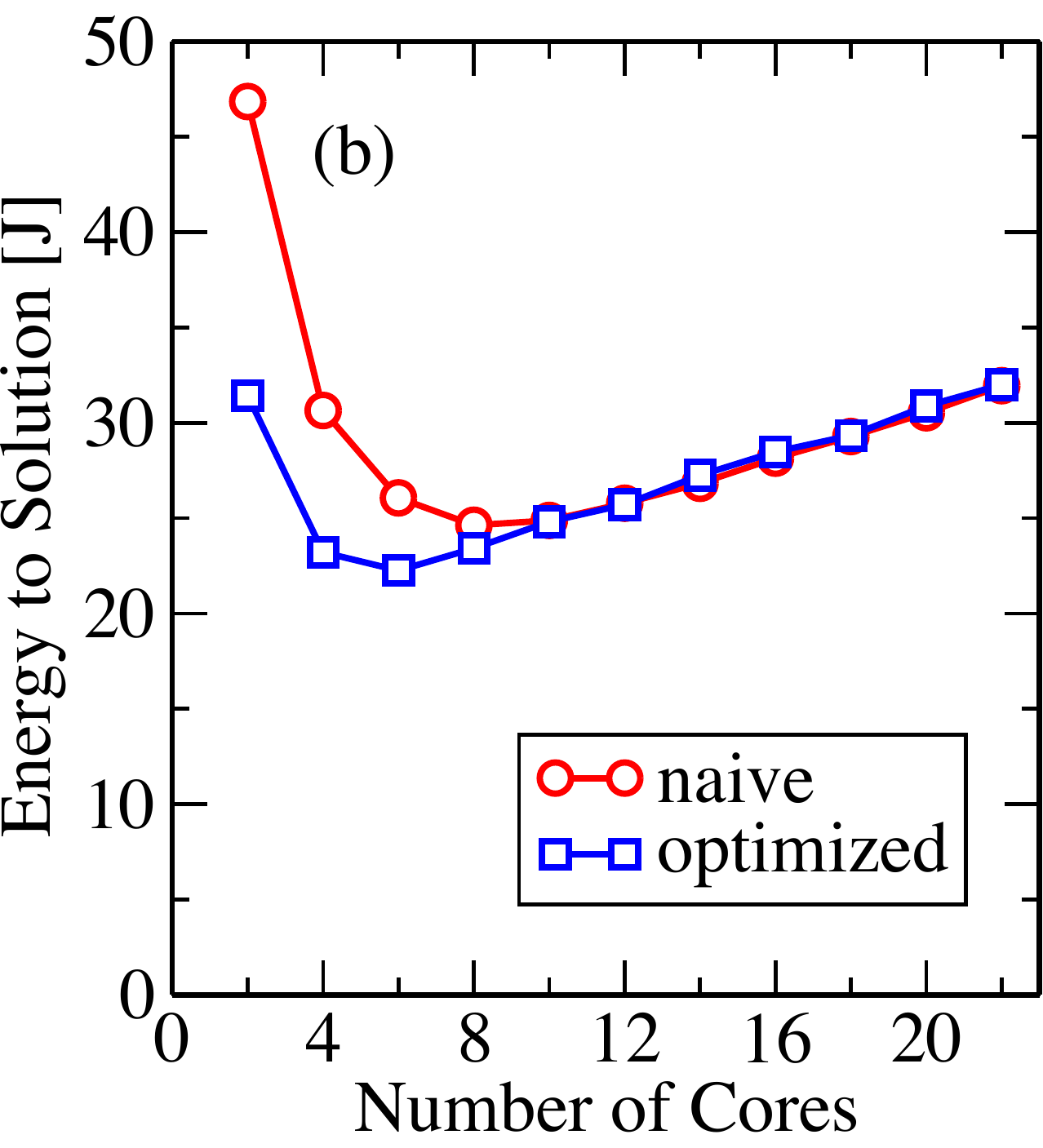}
    \caption{\label{fig:blueprint-example}Bandwidth and energy-to-solution for different 2D Jacobi implementations using a dataset size of 16\,GB.}
\end{figure}

This relation is illustrated in Fig.~\ref{fig:blueprint-example}, which depicts
measurement results obtained on a single socket of a standard two-socket
Broadwell-EP machine (cf. Section~\ref{sec:testbed} for hardware details).
Fig.~\ref{fig:blueprint-example}a shows memory bandwidth as function of active
cores for two different 2D Jacobi implementations.
Fig.~\ref{fig:blueprint-example}b depicts the energy-to-solution required for
both implementations.  From this real-world example two important conclusion
can be drawn:\\

\noindent 1. \emph{Memory-bound codes should be run using $n_\mathrm{s}$
cores}\\
Once the bandwidth bottleneck is hit, adding more cores no longer increases
performance.\footnote{In practise performance will decrease slightly when
using more than $n_\mathrm{s}$ cores, as prefetchers and memory subsystem
have to deal with additional memory streams which makes them less efficient;
this effect can be observed in Fig.~\ref{fig:blueprint-example}a.} Instead
using more cores increases chip power consumption resulting in a higher
energy-to-solution.  This effect is visible in
Fig.~\ref{fig:blueprint-example}b.\\

\noindent 2. \emph{Early saturation can lead to lower energy-to-solution}\\
The optimized version saturates memory bandwidth with six cores compared to the
ten cores required by the naive counterpart. Six vs. ten cores being active
typically translates into a lower power draw for version using fewer
cores.\footnote{In theory it is possible that an implementations using fewer
cores to saturate memory bandwidth is less energy-efficient than one that uses
more.  Consider, for example, an optimization that doubles single-core
performance but triples the power drawn by the core. $n_\mathrm{s}$ is halved
but energy to solution is 50\% higher nonetheless. In practise we have however
never observed such a scenario. For all optimizations described in
Section~\ref{sec:scopt} the increase in single-core power draw accompanying an
optimization is negligible.} Together with the fact that versions
($16\,\mathrm{GB}/b_\mathrm{s}$), this results in a better energy-to-solution
for the optimized version.

After establishing these facts we propose the following approach to
optimize energy consumption:
(1) Use the \ac{ECM} model to guide performance improvements for the
            single-core implementation.
(2) Attempt to lower per-core power draw by tuning hardware parameters,
            e.g. core frequency or \ac{COD} mode.
(3) Never run code with more cores than necessary ($n_\mathrm{s}$).


\begin{table*}[!tb]
\centering
\caption{\label{tab:testbed}Test machine specifications.}
\resizebox{\textwidth}{!}{%
\begin{tabular}{|l|c|c|c|c|}                                                                                                                                                                    \hline
Microarchitecture (Shorthand)   & Sandy Bridge-EP (SNB)             & Ivy Bridge-EP (IVB)                   & Haswell-EP (HSW)                      & Broadwell-EP  (BDW) \\                    \hline
Chip Model                      & Xeon E5-2680                      & Xeon E5-2690 v2                       & Xeon E5-2695 v3                       & Pre-release \\                            \hline
Release Date                    & Q1/2012                           & Q3/2013                               & Q3/2014                               & Q1/2016 \\                                \hline
non-AVX/AVX Base Freq.          & 2.7\,GHz/2.7\,GHz                 & 3.0\,GHz/3.0\,GHz                     & 2.3\,GHz/1.9\,GHz                     & 2.1\,GHz/2.0\,GHz \\                      \hline
Cores/Threads                   & 8/16                              & 10/20                                 & 14/28                                 & 18/36 \\                                  \hline
Latest SIMD Extensions          & AVX                               & AVX                                   & AVX2, FMA3                            & AVX2, FMA \\                              \hline
Core-Private L1/L2 Caches       & 8$\times$32\,kB/8$\times$256\,kB  & 10$\times$32\,kB/10$\times$256\,kB    & 14$\times$32\,kB/14$\times$256\,kB    & 18$\times$32\,kB/18$\times$256\,kB \\     \hline
Shared Last-Level Cache         & 20\,MB (8$\times$2.5\,MB)         & 25\,MB (10$\times$2.5\,MB)            & 35\,MB (14$\times$2.5\,MB)            & 45\,MB (18$\times$2.5\,MB) \\             \hline
Memory Configuration            & 4 ch. DDR3-1600                   & 4 ch. DDR3-1866                       & 4 ch. DDR4-2133                       & 4 ch. DDR4-2133 \\                        \hline
Theoretical Mem. Bandwidth      & 51.2\,GB/s                        & 59.7\,GB/s                            & 68.2\,GB/s                            & 68.2\,GB/s \\                             \hline
Sustained Copy Bandwidth        & 39.5\,GB/s (77\%)                 & 43.4\,GB/s (74\%)                     & 50.1\,GB/s (73\%)                     & 51.5 (76\%) \\                            \hline
L1$\rightarrow$Reg Bandwidth    & 2$\times$16\,B/cy                 & 2$\times$16\,B/cy                     & 2$\times$32\,B/cy                     & 2$\times$32\,B/cy \\                      \hline
Reg$\rightarrow$L1 Bandwidth    & 1$\times$16\,B/cy                 & 1$\times$16\,B/cy                     & 1$\times$32\,B/cy                     & 1$\times$32\,B/cy \\                      \hline
L1$\leftrightarrow$L2 Bandwidth & 32\,B/cy${\hat=}$2\,cy/CL         & 32\,B/cy${\hat=}$2\,cy/CL             & 64\,B/cy${\hat=}$1\,cy/CL             & 64\,B/cy${\hat=}$1\,cy/CL     \\          \hline
L2$\leftrightarrow$L3 Bandwidth & 32\,B/cy${\hat=}$2\,cy/CL         & 32\,B/cy${\hat=}$2\,cy/CL             & 32\,B/cy${\hat=}$2\,cy/CL             & 32\,B/cy${\hat=}$2\,cy/CL     \\          \hline
L3$\leftrightarrow$Mem Bandwidth (copy) & 14.5\,B/cy${\hat=}$4.4\,cy/CL & 20.0\,B/cy${\hat=}$4.4\,cy/CL     & 21.7\,B/cy${\hat=}$2.9\,cy/CL         & 24.6\,B/cy${\hat=}$2.6\,cy/CL \\          \hline
\end{tabular}}
\end{table*}

\section{The ECM Performance Model}
\label{sec:ecm}

The \ac{ECM} model \cite{Treibig:2009,hager:cpe,sthw15, Hofmann2016:1,
Hofmann2016:2} is an analytic performance model that, with the exception of
sustained memory bandwidth, works exclusively with architecture specifications
as inputs. The model estimates the numbers of CPU cycles required to execute a
number of iterations of a loop on a single core of a multi- or many-core chip.
For multi-core estimates, linear scaling of single-core performance is assumed
until a shared bottleneck, such as e.g.  main memory bandwidth, is hit.  Note
that only parts of the model relevant to this work, i.e. the single-core model,
are presented here.  Readers interested in the full ECM model can find the most
recent version for Intel Xeon, Intel Xeon Phi, and IBM POWER8 processors here
\cite{CPE:CPE3921}.


The single-core prediction is made up of contributions from the in-core
execution time $T_\mathrm{core}$, i.e. the time spent executing instructions in
the core under the assumption that all data resides in the L1 cache, and the
transfer time $T_\mathrm{data}$, i.e. the time spent transferring data from its
location in the cache/memory hierarchy to the L1 cache.  As data transfers in
the cache- and memory hierarchy occur at \ac{CL} granularity we chose the
number of loop iterations $n_\mathrm{it}$ to correspond to one cache line's
"worth of work." On Intel architectures, where \ac{CL}s are 64\,B in size,
$n_\mathrm{it}=8$ when using \ac{DP} floating-point numbers, because processing
eight ``doubles'' (8\,B each) corresponds to exactly one \ac{CL} worth of work.

Superscalar core designs house multiple execution units, each dedicated to
perform certain work: loading, storing, multiplying, adding, etc.  The in-core
execution time $T_\mathrm{core}$ is determined by the unit that takes the
longest to retire the instructions allocated to it. Other constraints for the
in-core execution time may apply, e.g. the four micro-op per cycle retirement
limit of Intel Xeon cores. The model differentiates between core cycles
depending on whether data transfers in the cache hierarchy can overlap with
in-core execution time. For instance, on Intel Xeons, core cycles in which data
is moved between the L1 cache and registers, e.g. cycles in which load and/or
store instructions are retired, prohibit simultaneous transfer of data between
the L1 and L2 cache; these ``non-overlapping'' cycles contribute to
$T_\mathrm{nOL}$. Cycles in which no load or store instructions but other
instructions, such as e.g.  arithmetic instructions, retire are considered
``overlapping'' cycles and contribute to $T_\mathrm{OL}$. The in-core runtime
is the maximum of both: $T_\mathrm{core} = \max(T_\mathrm{OL},
T_\mathrm{nOL})$.

For modelling data transfers, latency effects are initially neglected, so
transfer times are exclusively a function of bandwidth. Cache bandwidths are
typically well documented and can be found in vendor data sheets. Depending on
how many \ac{CL}s have to be transferred, the contribution of each level in the
memory hierarchy ($T_\mathrm{L1L2}$, \ldots, $T_\mathrm{L3Mem}$) can be
determined.  Special care has to be taken when dealing with main memory
bandwidth, because theoretical memory bandwidth specified in the data sheet and
sustained memory bandwidth $b_\mathrm{s}$ can differ greatly. Also, in
practise $b_\mathrm{s}$ depends on the number of load and store
streams. It is therefore recommended to empirically determine $b_\mathrm{s}$
using a kernel that resembles the memory access pattern of the benchmark to be
modeled.  Once determined, the time to transfer one
\ac{CL} between the L3 cache and main memory can be derived from the CPU
frequency $f$ as $64\,\mathrm{B} \cdot f / b_\mathrm{s}$~cycles.

Starting with the Haswell-EP (HSW) microarchitecture, an empirically determined
latency penalty $T_p$ is applied to off-core transfer times.  This departure
from the bandwidth-only model has been made necessary by large core counts, the
dual-ring design, and separate clock frequencies for core(s) and Uncore all of
which increase latencies when accessing off-core data.  The penalty is added
each time the Uncore interconnect is involved in data transfers. This is the
case whenever data is transferred between the L2 and L3 caches, as data is
pseudo-randomly distributed between all last-level cache segments; and when
data is transferred between the L3 cache and memory, because the memory
controller is attached to the Uncore interconnect.  Instruction times as well
as data transfer times, e.g.  $T_\mathrm{L1L2}$ for the time required to
transfer data between L1 and L2 caches, can be summarized in shorthand
notation:
\ecmshort{T_\mathrm{OL}}{T_\mathrm{nOL}}{T_\mathrm{L1L2}}{T_\mathrm{L2L3}+T_\mathrm{p}}{T_\mathrm{L3Mem}+T_\mathrm{p}}.

To arrive at a prediction, in-core execution and data transfers times are put
together. Depending on whether there exist enough overlapping cycles to hide
all data transfers, runtime is given by either $T_\mathrm{OL}$ or the sum of
non-overlapping core cycles $T_\mathrm{nOL}$ plus contributions of data
transfers $T_\mathrm{data}$, whichever takes longer. $T_\mathrm{data}$ consists
of all necessary data transfers in the cache/memory hierarchy, plus latency
penalties if applicable, e.g. for data coming from the L3 cache:
$T_\mathrm{data}=T_\mathrm{L1L2}+T_\mathrm{L2L3}+T_\mathrm{p}$.  The prediction
is thus $T_\mathrm{ECM} = \max(T_\mathrm{OL}, T_\mathrm{nOL} +
T_\mathrm{data})$. A shorthand notation also exists for the model's prediction:
\ecmpshort{T_\mathrm{ECM}^\mathrm{core}}{T_\mathrm{ECM}^\mathrm{L2}}{T_\mathrm{ECM}^\mathrm{L3}}{T_\mathrm{ECM}^\mathrm{Mem}}{}.

Converting the prediction from time (in cycles) to performance (work per
second) is done by dividing the work per CL $W_\mathrm{CL}$ (e.g.
floating-point operations, updates, or any other relevant work metric) by the
predicted runtime in cycles and multiplying with the processor frequency $f$,
i.e.  $P_\mathrm{ECM} = W_\mathrm{CL}/T_\mathrm{ECM} \cdot f$.

%
%
%


\section{Experimental Testbed}
\label{sec:testbed}

All measurements were performed on standard two-socket Intel Xeon servers.  A
summary of key specifications of the four generations of processors can be
found in Table~\ref{tab:testbed}.  According to Intel's ``tick-tock'' model,
where a ``tick'' corresponds to a shrink of the manufacturing process
technology and a ``tock'' to a new microarchitecture, IVB and BDW are
``ticks''---apart from an increase in core count and a faster memory clock, no
major improvements were introduced in these microarchitectures.

HSW, which is a ``tock'', introduced AVX2, extending the already existing
256\,bit SIMD vectorization from floating-point to integer data types.
Instructions introduced by the \ac{FMA} extension are handled by two new,
AVX-capable execution units. Data paths between the L1 cache and registers as
well as the L1 and L2 caches were doubled. Due to limited scalability of a
single ring connecting the cores, HSW chips with more than eight feature a
dual-ring design.  HSW also introduces the AVX base and maximum AVX Turbo
frequencies.  The former is the minimum guaranteed frequency when running AVX
code on all cores; the latter the maximum frequency when running AVX code on
all cores (cf.  Table~3 in \cite{Intel-v3-specupdate}).  Based on workload, the
actual frequency varies between this minimum and maximum value. For a more
detailed analysis of the differences between the SNB/IVB and HSW
microarchitectures see \cite{Hofmann2016:1}.

For all measurements on SNB and IVB, the CPU frequency was fixed at the nominal
CPU frequency.  On HSW and BDW, the CPU frequency was fixed to the non-AVX base
frequency;\footnote{The guaranteed frequency when running code on all cores
that does not use AVX instructions.} in none of the single-core measurements
that feature AVX code could we observe a drop below the non-AVX base frequency.
The measured sustained main memory bandwidth displayed in
Table~\ref{tab:testbed} is that of the STREAM copy kernel \cite{McCalpin1995}
using non-temporal stores, because its memory access pattern corresponds to
that of the 2D Jacobi solver.  Energy consumption was determined by accessing
the \ac{RAPL} interface through \texttt{likwid-perfctr} \cite{Treibig:2012}.
Because node and memory power usage fluctuates depending on node configuration
and RAM manufacturer, we chose to only present results pertaining to chip power
consumption, i.e. cores, core-private caches, and Uncore (cf.~14.9 in
\cite{intel-sdm-2016}).  All code is compiled using the Intel C Compiler
version 15.0.2.


\section{Single-Core Optimizations}
\label{sec:scopt}

\begin{figure}[tb]
\lstset{
        breaklines=true,
        language=C,
        basicstyle=\small\ttfamily,
        numbers=left,
        numberstyle=\tiny,
        frame=t,
        columns=fullflexible,
        showstringspaces=false,
        escapechar=\%,
        escapebegin=\color{blue}\ttfamily\bfseries,
        keepspaces=true
}
\begin{lstlisting}
for (y=1; y<Y-1; ++y)
  for (x=1; x<X-1; ++x)
    b[y][x]=0.25 * (a[y-1][x] + a[y][x-1] +
                    a[y][x+1] + a[y+1][x]);
\end{lstlisting}
\caption{\label{fig:code:baseline} C implementation for one 2D Jacobi iteration.}
\end{figure}

Figure~\ref{fig:code:baseline} shows the source code for one 2D five-point
Jacobi sweep, i.e. the complete update of all grid points.  One grid point
update computes and stores in \texttt{b} the new state of each point from the
values of its four neighbors in \texttt{a}, which holds data from the previous
iteration.  For results to be representative, the dataset size per socket for
all measurements is 16\,GB, i.e. each of the two grids is 8\,GB in size and
made up of $32768\times32768$ double-precision numbers.  Performance is
expressed in ``lattice updates per second'' (LUP/s), i.e. scalar inner kernel
iterations per second.

\subsection{Baseline AVX Implementation}
\label{sec:scopt:baseline}

Using adequate optimization flags (\texttt{-O3 -xHost -fno-alias}) it is
trivial to generate AVX vectorized assembly for the code shown in
Figure~\ref{fig:code:baseline} with recent Intel compilers. This is why we
decided to use an AVX vectorized variant instead of scalar code as baseline.

With 256-bit AVX vectorization in place, one \ac{CL} worth of work (eight
LUP/s), consists of eight AVX loads, two AVX stores, six AVX adds, and two AVX
multiplication instructions.

To determine data transfers inside the cache hierarchy, we have to examine each
load and store in more detail. Storing the newly computed results to array
\texttt{b} involves the transfer of two \ac{CL}s to/from main memory: because
both arrays are too large to fit inside the caches, the store will miss in the
L1 cache, triggering a write-allocate of the \ac{CL} from main memory. After
the values in the \ac{CL} have been updated, the \ac{CL} will have to be
evicted from the caches eventually, triggering another main memory transfer.

The left neighbor \texttt{a[y][x-1]} can always be loaded from the L1 cache
since it was used two inner iterations before as right neighbor; access to
\texttt{a[y+1][x]} must be loaded from main memory since it was not used before
within the sweep.

Based on work by Rivera and Tseng \cite{Rivera:2000} Stengel et al. introduced
the layer condition (LC) \cite{sthw15} to help determining where data for
\texttt{a[y-1][x]} and \texttt{a[y][x+1]} is coming from.  The LC stipulates
three successive rows have to fit into a certain cache for accesses to
these data to come from this particular cache.
Assuming cache $k$ can effectively hold data up to 50\% of its nominal size
$C_k$, the LC can be formulated as
$3 \cdot N \cdot 8\,\mathrm{B} < 50\% \cdot C_k$.
For $N=32768$, three rows take up 768\,kB so on all previously introduced
machines (cf. line~8 in Table~\ref{tab:testbed}) the LC holds true for the L3
cache.\\

\noindent\textbf{ECM Model for SNB and IVB}
\label{sec:scopt:baseline:SNBIVB}

To process one CL worth of data, eight AVX load, two AVX store, six AVX
addition and two AVX multiplication instructions have to executed.  Throughput
is limited by the two load units. Each load unit has a 16\,B wide data path
connecting registers and L1 cache, so retiring a 32\,B AVX load takes two
cycles. Using both load units, eight AVX loads take
$T_\mathrm{nOL}=8\,\mathrm{cy}$. Both AVX stores are retired in parallel with the
eight loads, so they do not increase $T_\mathrm{nOL}$.  Computation throughput
is limited by the single add port, which takes $T_\mathrm{OL}=6\,\mathrm{cy}$
to retire all six AVX add instructions. Both AVX multiplications can be
processed in parallel with two of the six AVX add instructions.

As established previously, three CLs have to be transferred between L3 and
memory: write-allocating and later evicting \texttt{b[y][x]} and loading
\texttt{a[y+1][x]}. On both SNB and IVB, L3-memory bandwidth is 4.4\,cy/CL (cf.
last line in Table~\ref{tab:testbed}). This results in
$T_\mathrm{L3Mem}=13.2\,\mathrm{cy}$ for both architectures.  The same three
CLs have to be transferred between L3 and L2 cache; in addition, CLs for
\texttt{a[y-1][x]} and \texttt{a[y][x+1]} have to be transferred from the L3 to
the L2 cache. Transferring five CLs at a bandwidth of 2\,cy/CL takes
$T_\mathrm{L2L3}=10\,\mathrm{cy}$ on both SNB and IVB. At a L1-L2 bandwidth of
2\,cy/CL, moving these five CLs between L2 and L1 cache takes
$T_\mathrm{L1L2}=10\,\mathrm{cy}$.  Using the ECM short notation to summarize
the inputs yields \ecm{6}{8}{10}{10}{13.2}{\mathrm{cy}} for both SNB and IVB;
the corresponding runtime prediction for SNB and IVB is
\ecmp{8}{18}{28}{41.2}{\mathrm{cy}}.

For a 2.7\,GHz SNB core the performance prediction is $P_\mathrm{ECM}^{Mem}=
\frac{8\,\mathrm{LUP/CL}}{41.2\,\mathrm{cy/CL}} \cdot 2.7\,\mathrm{GHz} =
524\,\mathrm{MLUP/s.}$ For IVB the model
predicts a performance of $582\,\mathrm{MLUP/s}$.\\


\noindent\textbf{ECM Model for HSW and BDW}
\label{sec:scopt:baseline:HSWBDW}

On HSW and BDW, the \acp{AGU} are the bottleneck for $T_\mathrm{nOL}$.  Each
load/store instruction accesses an \ac{AGU} to compute the referenced memory
address. With only two \acp{AGU} capable of performing the required addressing
operations available, retiring all ten load/store instructions takes
$T_\mathrm{nOL}=5\,\mathrm{cy}$.  HSW and BDW posses a single AVX add unit just
like SNB and IVB, so $T_\mathrm{OL}=6\,\mathrm{cy}$.

The off-core latency penalty was empirically estimated at approximately
1.6~cycles for both HSW and BDW and is applied per CL transfer that takes place
over the Uncore interconnect, i.e. all transfers between L3 and L2 caches as
well as transfers between memory and the L3 cache. The effective L3-Mem
bandwidth is thus 2.9+1.6\,cy on HSW and 2.6+1.6\,cy on BDW; the effective
L2-L3 is 2+1.6\,cy on HSW and 2+1.6\,cy on BDW.

Transferring the three required CLs then results in
$T_\mathrm{L3Mem}=8.7+4.8\,\mathrm{cy}$ on HSW resp.
$T_\mathrm{L3Mem}=7.8+4.8\,\mathrm{cy}$ on BDW. Moving five CLs between the L2
and L3 caches takes $T_\mathrm{L2L3}=10+8\,\mathrm{cy}$ on both HSW and BDW. At
a L1-L2 bandwidth of 1\,cy/CL, moving the same five CLs takes
$T_\mathrm{L1L2}=5\,\mathrm{cy}$ on both microarchitectures.  The ECM inputs
are thus \ecm{6}{5}{5}{10+8}{8.7+4.8}{\mathrm{cy}} for HSW and for BDW
\ecm{6}{5}{5}{10+8}{7.8+4.8}{\mathrm{cy}}. The corresponding runtime
predictions are \ecmp{6}{10}{28}{41.5}{\mathrm{cy}} for HSW and
\ecmp{6}{10}{28}{40.6}{\mathrm{cy}} for BDW. The performance predictions are
443\,MLUP/s for HSW and 413\,MLUP/s for BDW.


\subsection{Cache Blocking Optimization}
\label{sec:scopt:cacheblocking}

One way to increase the performance of the single-core implementation is to
reduce the amount of time spent transferring data inside the cache hierarchy.
As previously established, \texttt{a[y-1][x]} and \texttt{a[y][x+1]} are
loaded from the L3 cache, because the L1 and L2 caches are too small to
fulfill the LC for $N=32768$. Using cache blocking, it is possible to
enforce the LC in arbitrary cache levels.  This is done by partitioning the
grid into stripes along the $y$-axis; the grid is then processed stripe by
stripe.  The diameter of the stripes, also known as blocking factor $b_x$, is
chosen is such a way that the LC is met for a given cache level.
If L2 blocking is desired, the L2 cache size of 256\,kB requires that $b_x$
should be chosen smaller than 5461.

Efficient blocking for the 32\,kB L1 cache not as straightforward. Although
determining $b_x<682$ is simple using the LC, naive L1 blocking
in $x$-direction has negative side effects. With most of the data for one CL
update coming from L1, the L2 cache is less busy. This slack is detected by
hardware prefetchers, making them more aggressive, leading to data being
prefetched from main memory. With $b_x \approx 680$, the size of one stripe is
$680 \cdot 32768 \cdot 8\,\mathrm{B}=170\,\mathrm{MB}$---too large for the L3
cache, which means that data prefetched from main memory will be preempted from
the cache before it is used. This can be avoided either by disabling some of
the prefetchers or additionally blocking in $y$-direction. We used the latter,
because disabling prefetchers might degrade performance elsewhere.  To
guarantee data is used before being preempted, the size of each
chunk should be chosen smaller than 50\% of a single L3 segment,\footnote{In
the multi-core scenario, all cores might be active and store data in the shared
L3 cache; the capacity dedicated to a single core is thus that of a single L3
segment adjusted by 50\% to reflect effective cache size.} e.g. $b_y < 50\%
\cdot 2.5\,\mathrm{MB} / ( b_x \cdot 8\,\mathrm{B})$.\\

\noindent\textbf{ECM Model for SNB and IVB}
\label{sec:scopt:cacheblocking:SNBIVB}

Other than causing negligible loop overhead cache blocking does not change the
instructions that have to be retired to process one CL; thus $T_\mathrm{OL}$
and $T_\mathrm{nOL}$ remain unchanged.

L2 blocking reduces the number of CLs transferred between L3 and L2 from five
to three, lowering $T_\mathrm{L2L3}$ from ten to six cycles on both SNB and
IVB. The runtime prediction $T_\mathrm{ECM}^\mathrm{Mem}$ for both SNB
and IVB is reduced from 41.2 to 37.2\,cy, leading to a performance prediction
$P_\mathrm{ECM}^\mathrm{Mem}$ of 580\,MLUP/s for SNB and 645\,MLUP/s for IVB.

Similarly L1 blocking reduces the CL traffic between L2 and L1 caches from five
to three CLs, lowering $T_\mathrm{L1L2}$ from ten to six cycles on both SNB and
IVB. Again, the runtime prediction $T_\mathrm{ECM}^\mathrm{Mem}$ for both
microarchitectures is reduced by four cycles from 37.2 to 33.2\,cy.  The
predicted performance $P_\mathrm{ECM}^\mathrm{Mem}$ increases to 651\,MLUP/s on
SNB and 723\,MLUP/s on IVB.\\


\noindent\textbf{ECM Model for HSW and BDW}
\label{sec:scopt:cacheblocking:HSWBDW}

The effect of L2 blocking is more pronounced on HSW and BDW, because the cost
of transferring a CL is higher on these microarchitectures due to the latency
penalty. L2 blocking lowers $T_\mathrm{L2L3}$ from 10+8 to 6+4.8\,cy on HSW and
BDW. In turn, the runtime prediction $T_\mathrm{ECM}^\mathrm{Mem}$ for HSW is
reduced from 41.5 to 34.3\,cy and from 40.6 to 33.4\,cy on BDW.  The
performance prediction $P_\mathrm{ECM}^\mathrm{Mem}$ increases to 536\,MLUP/s
on HSW and 503\,MLUP/s on BDW.

Because the L1-L2 bandwidth increased from 2\,cy/CL on SNB/IVB to 1\,cy/CL on
HSW/BDW, the performance improvement offered by L1 blocking is less pronounced
than on SNB and IVB. With the number of CL transfers lowered from five to three
with L1 blocking, $T_\mathrm{L1L2}$ is reduced from five to three cycles.  The
runtime prediction $T_\mathrm{ECM}^\mathrm{Mem}$ for HSW is reduced from 34.3
to 32.3\,cy; on BDW from 33.4 to 31.4\,cy.  The predicted performance
$P_\mathrm{ECM}^\mathrm{Mem}$ is increased to 570\,MLUP/s on HSW and
535\,MLUP/s on BDW.



\subsection{Cluster-on-Die Mode}
\label{sec:scopt:cod}

As a workaround to the limited scalability of the physical ring interconnect
introduced with Westmere-EX, HSW and BDW switch to a dual-ring design. HSW uses
the so-called ``eight plus $x$'' design, in which the first eight cores of a
chip are attached to a primary ring and the remaining cores (six for the
model introduced in Section~\ref{sec:testbed}) are attached to a secondary
ring; BDW uses a symmetric design. Two queues enable data to pass between
rings. In the default (non-\ac{COD}) mode, the physical topology is hidden from the
operating system, i.e. all cores are exposed within the same \ac{NUMA} domain.

To understand the latency problems caused by the interconnect, we examine the
route data travels inside the Uncore.  Using a hashing function data is is
distributed across all L3 segments based on its memory address.  When accessing
data in the L3 cache, there is a high probability it must be fetched from
remote L3 segments.  In the worst case, this is a segment on the other ring so
there might be a high latency involved. In the case of a L3 miss the situation
gets worse. Each physical ring has attached to it a \ac{MC} and the
choice which MC to use is again based on the data's address. So a L3 miss in a
segment on one physical ring does not imply that this ring's \ac{MC} will be
used to fetch the data from memory.  That leads to cases in which a large
number of hops and multiple cross-physical ring transfers are involved when
getting data from main memory.

One way to reduce these latencies is the new \ac{COD} mode introduced together
with the dual-ring design on HSW and BDW in which cores are separated into
two physical clusters of equal size.
The latency reduction is achieved by adapting the involved hashing functions.
Data requested by a core of a cluster will only be placed in the cluster's L3
segments; in addition, all memory transfers are routed to the \ac{MC} dedicated
to the cluster.  Thus, for \ac{NUMA}-aware codes, \ac{COD} mode effectively
lowers the latency by reducing the diameter and the mean distance of the
dual-ring interconnect by restricting each cluster to its own physical ring.
For a more detailed analysis of \ac{COD} mode see \cite{Hofmann2016:1}. On the
HSW chip used for benchmarks, \ac{COD} mode lowers the interconnect latency by
0.5\,cy; on the employed BDW chip, where a single ring still has eleven cores
attached to it, the latency is only reduced by 0.3\,cy.\\

\noindent\textbf{ECM Model for HSW and BDW}
\label{sec:scopt:cod:HSWBDW}

With \ac{COD} enabled, the per-CL Uncore latency penalty is reduced to 1.1\,cy on HSW resp.
1.3\,cy on BDW.  This leads to $T_\mathrm{L3Mem}=8.7+3.3$\,cy on HSW resp.
7.8+3.9\,cy on BDW.  Because the Uncore is also involved in L2-L3 transfers,
the latency improvement caused by enabling \ac{COD} also positively influences
$T_\mathrm{L2L3}$. To transfer three CLs, 6+3.3\,cy are required on HSW resp.
6+3.9\,cy on BDW.  The resulting ECM inputs are \ecmshort{6}{5}{3}{6+3.3}{8.7+3.3}
for HSW and \ecmshort{6}{5}{3}{6+3.9}{7.8+3.9} for BDW.  The corresponding
runtime prediction is \ecmpshort{6}{8}{17.3}{29.3} on HSW and
\ecmpshort{6}{8}{17.9}{29.6} on BDW. The performance predicted by the ECM model
is 628\,MLUP/s for HSW and 567\,MLUP/s on BDW.


\subsection{Non-Temporal Stores}
\label{sec:scopt:ntstores}

Streaming or \ac{NT} stores are special instructions that
avoid write-allocates on modern Intel microarchitectures.
Without NT stores, storing the newly computed result \texttt{b[y][x]} triggers
a write-allocate. The old data is brought in from memory and travels through
the whole cache hierarchy.  Thus the first benefit of using NT stores is that
the unnecessary transfer of \texttt{b[y][x]} from memory to the L1 cache no
longer takes place. In addition, NT stores will also strip some cycles off the
time involved getting the new result to main memory. Using regular stores, the
newly computed result is written to the L1 cache, from where the data has to be
evicted at some point through the whole cache hierarchy into memory.  Using NT
stores, \acp{CL} are sent via the L1 cache to the \acp{LFB}; from there, they
are transfered directly to memory and do not pass through the L2 and L3 caches.
Although the benefits should apply equally to all microarchitectures, there are
shortcomings in SNB and IVB that make single-core implementations using NT
stores slower than their regular stores counterpart. The positive impact of NT
stores can only be leveraged in multi-core scenarios on these
microarchitectures, which is why we chose to omit ECM models and measurements
for the NT store implementation for SNB and IVB.\\

\begin{table*}[!tb]
\centering
\caption{\label{tab:single-core-results}Summary of ECM inputs, ECM predictions, as well as measured performance, power consumption, and energy-to-solution for one 2D Jacobi iteration with a dataset size of 16\,GB.}
\resizebox{\textwidth}{!}{%
\begin{tabular}{|l|l|c|c|c|c|c|c|}                                                                                                                                                                                                      \hline
                &               & ECM                                                           & ECM pre-                                                  & $P_\mathrm{ECM}^\mathrm{Mem}$ & Measured  & Chip      & Chip Energy-\\    
\textmu{}arch   & Version       & input [cy]                                                    & diction [cy]                                              & [MLUP/s]      & [MLUP/s]  & Power [W] & to-Solution [J] \\                \hline
SNB             & Baseline      & \ecmshort{6}{8}{10}{10}{13.2}                                 & \ecmpshort{8}{18}{28}{41.2}                               & 524           & 514       & 35.9      & 75.0 \\                           
                & L2 blocked    & \ecmshort{6}{8}{10}{\textbf{6}}{13.2}                         & \ecmpshort{8}{18}{\textbf{24}}{\textbf{37.2}}             & 580           & 623       & 36.4      & 62.7 \\                           
                & L1 blocked    & \ecmshort{6}{8}{\textbf{6}}{6}{13.2}                          & \ecmpshort{8}{\textbf{14}}{\textbf{20}}{\textbf{33.2}}    & 651           & 672       & 36.6      & 58.4 \\                           \hline
IVB             & Baseline      & \ecmshort{6}{8}{10}{10}{13.2}                                 & \ecmpshort{8}{18}{28}{41.2}                               & 582           & 539       & 32.7      & 65.3 \\                           
                & L2 blocked    & \ecmshort{6}{8}{10}{\textbf{6}}{13.2}                         & \ecmpshort{8}{18}{\textbf{24}}{\textbf{37.2}}             & 645           & 651       & 34.1      & 56.1 \\                           
                & L1 blocked    & \ecmshort{6}{8}{\textbf{6}}{6}{13.2}                          & \ecmpshort{8}{\textbf{14}}{\textbf{20}}{\textbf{33.2}}    & 722           & 714       & 33.0      & 49.6 \\                           \hline
HSW             & Baseline      & \ecmshort{6}{5}{5}{10+8}{8.7+4.8}                             & \ecmpshort{6}{10}{28}{41.5}                               & 443           & 435       & 50.9      & 125.6\\                           
                & L2 blocked    & \ecmshort{6}{5}{5}{\textbf{6+4.8}}{8.7+4.8}                   & \ecmpshort{6}{10}{\textbf{20.8}}{\textbf{34.3}}           & 536           & 529       & 51.0      & 103.5\\                           
                & L1 blocked    & \ecmshort{6}{5}{\textbf{3}}{6+4.8}{8.7+4.8}                   & \ecmpshort{6}{\textbf{8}}{\textbf{18.8}}{\textbf{32.3}}   & 570           & 579       & 52.1      & 96.6 \\                           
                & L1 b.+CoD     & \ecmshort{6}{5}{3}{6+\textbf{3.3}}{8.7+\textbf{3.3}}          & \ecmpshort{6}{8}{\textbf{17.3}}{\textbf{29.3}}            & 628           & 625       & 51.3      & 88.0 \\                           
                & L1 b.+CoD+nt  & \ecmshort{6}{5}{\textbf{2}}{\textbf{2+1.1}}{\textbf{5.8+2.2}} & \ecmpshort{6}{\textbf{7}}{\textbf{10.1}}{\textbf{18.1}}   & 1016          & 951       & 51.0      & 57.6 \\                           \hline
BDW             & Baseline      & \ecmshort{6}{5}{5}{10+8}{7.8+4.8}                             & \ecmpshort{6}{10}{28}{40.6}                               & 413           & 407       & 44.2      & 116.5\\                           
                & L2 blocked    & \ecmshort{6}{5}{5}{\textbf{6+4.8}}{7.8+4.8}                   & \ecmpshort{6}{10}{\textbf{20.8}}{\textbf{33.4}}           & 503           & 489       & 44.3      & 97.4 \\                           
                & L1 blocked    & \ecmshort{6}{5}{\textbf{3}}{6+4.8}{7.8+4.8}                   & \ecmpshort{6}{\textbf{8}}{\textbf{18.8}}{\textbf{31.4}}   & 535           & 509       & 44.5      & 93.7 \\                           
                & L1 b.+CoD     & \ecmshort{6}{5}{3}{6+\textbf{3.9}}{7.8+\textbf{3.9}}          & \ecmpshort{6}{8}{\textbf{17.9}}{\textbf{29.6}}            & 567           & 561       & 44.6      & 86.0 \\                           
                & L1 b.+CoD+nt  & \ecmshort{6}{5}{\textbf{2}}{\textbf{2+1.3}}{\textbf{5.2+2.6}} & \ecmpshort{6}{\textbf{7}}{\textbf{10.3}}{\textbf{18.1}}   & 928           & 862       & 45.4      & 56.6 \\                           \hline
\end{tabular}}
\end{table*}

\noindent\textbf{ECM Model for HSW and BDW}
\label{sec:scopt:ntstores:HSWBDW}

Transferring \texttt{a[y+1][x]} between the L1 and L2 caches takes 1\,cy.
Although the transfer is not strictly between the L1 and L2 cache, the cycle
spent transferring the CL for \texttt{b[y][x]} from the L1 cache to the
\acf{LFB} is booked in $T_\mathrm{L1L2}$ as well, making for a total L1-L2
transfer time of 2\,cy.  Loading \texttt{a[y+1][x]} from the L3 to the L2 cache
takes $T_\mathrm{L2L3}=2+1.1$\,cy on HSW and 2+1.3\,cy on BDW. Getting the CL
containing \texttt{a[y+1][x]} from memory takes 2.9+1.1\,cy on HSW and
2.6+1.3\,cy on BDW; again, although the transfer of \texttt{b[y][x]} is
strictly not between the L3 cache and memory, the transfer time to send the CL
from the \ac{LFB} to memory is booked in $T_\mathrm{L3Mem}$, making for a total
L3-Mem transfer time of 5.8+2.2\,cy on HSW and 5.2+2.6\,cy on BDW.  In summary,
the full ECM inputs are \ecmshort{6}{5}{2}{2+1.1}{5.8+2.2}\,cy on HSW and
\ecmshort{6}{5}{2}{2+1.3}{5.2+2.6}\,cy on BDW; the corresponding ECM runtime
prediction is \ecmpshort{6}{7}{10.1}{18.1} on HSW and
\ecmpshort{6}{7}{10.3}{18.1}.  The in-memory performance prediction by the ECM
model is $P_\mathrm{ECM}^\mathrm{Mem}=1016\,\mathrm{MLUP/s}$ on HSW and
928\,MLUP/s on BDW.



\section{Single-Core Results}
\label{sec:res:sc}

Table~\ref{tab:single-core-results} contains a summary of the ECM inputs and
predictions discussed in Section~\ref{sec:scopt}, as well as measurements of
performance, power, and energy consumption for one 2D Jacobi iteration using a
16\,GB dataset.  The model correctly predicts performance with a mean error of
3\% and a maximum error of 7\%, which indicates that all single-core
performance engineering measures work as intended. On SNB and IVB performance
increases of around 1.3$\times$ are achieved; on HSW with 2.2$\times$ resp. BDW
with 2.1$\times$ improvement the increases are even more pronounced.
Measurements obtained via the \ac{RAPL} interface indicate increases in power
consumption due to optimizations are negligible (in the range of 2\%). With
power draw almost constant, this means that the performance gains directly
translate into energy improvements.

An interesting observation regarding single-core power consumption\footnote{The
\ac{RAPL} counters can not report power consumption of individual cores but
only that of the whole package. Thus reported values also include power drawn
by Uncore facilities, e.g. all L3 segments and the interconnect.} can be made
when comparing the different microarchitectures that can be explained using
Intel's ``tick-tock'' model.  Power decreases with ``ticks,'' i.e. a shrink in
manufacturing size and the accompanying decreases in dynamic power; power
increases with ``tocks,'' i.e.  major improvements in microarchitecture: The
``tick'' from SNB which uses 32\,nm to IVB which uses 22\,nm technology
corresponds to a 10\% decrease in power consumption. HSW, the only ``tock'' in
Table~\ref{tab:single-core-results} uses the same 22\,nm process as IVB but
introduced major improvements in the microarchitecture (cf.
Section~\ref{sec:testbed}) which lead to a more than 50\% higher power draw.
BDW, a ``tick'' which uses a
14\,nm process uses 11\% less power than HSW.  While it is tempting to
generalize from these results, note that the reported numbers are specific to the
2D Jacobi application and the used chip models. We also believe that while a
surge in power consumption occurs with the HSW ``tock'' the cause to be a
combination of microarchitectural improvements and the higher core
count.\footnote{Despite only one core being active in the measurements, all L3
cache segments are active and draw power; a chip with more cores will thus draw
more power in single-core use cases.}


\section{Chip-Level Observations}
\label{sec:res:mc}




Although the \ac{ECM} model can be used to predict multi-core scaling behavior
\cite{CPE:CPE3921,Hofmann2016:1,sthw15}, due to space constraints chip-level discussion is restricted
to empirical results.
The graphs in Figure~\ref{fig:chip-perf} show the improvements
discussed in Section~\ref{sec:scopt} and relate measured performance and
energy-to-solution for different core counts.\footnote{The
leftmost measuring point of each graph corresponds to one core. Following the
line attached to a point to the next corresponds to one more core being active.
For demonstration purposes the purple graph in Fig.~\ref{fig:chip-perf}a has
some core counts annotated.} IVB (Fig.~{\ref{fig:chip-perf}a) and
HSW (Fig.~\ref{fig:chip-perf}b) are chosen as representatives to demonstrate
that the different optimizations have different impacts on performance and
energy consumption depending on the microarchitecture.

On IVB, running on $n_\mathrm{s} = 5$ instead of ten cores reduces energy
consumption by 13\% 44.4\,J to 34.8\,J; on HSW, running on seven instead of 14
cores amounts to a reduction of 16\% from 57.0\,J to 47.6\,J. The positive
impacts of L1 blocking can be observed for both microarchitectures,
corresponding to a further decrease of energy consumption by 16\% to 29.1\,J on
IVB resp. 10\% to 43.0\,J on HSW.  \ac{COD} mode 
brings energy consumption on HSW further down by 6\% to 40.5\,J.

When it comes to NT stores the differences in the microarchitectures
become visible. On IVB, the per-core performance of the implementation using
L1 blocking and NT stores (dark blue line in
Fig.~\ref{fig:chip-perf}a) is \textit{lower} than that of the version using
regular stores.  A version using NT stores without L1 blocking
(bright blue line) does not even manage to sustain memory bandwidth. Due to this bad
per-core performance, NT stores on SNB/IVB have almost no
positive impact on energy consumption!  On HSW, per-core performance as
expected is exactly $1.5\times$ faster with NT stores, enabling a
lowering in energy consumption by 33\% to 27.0\,J.

Another difference in microarchitecture surfaces when examining
frequency-tuning for potential energy savings.  Before HSW, i.e. on SNB and
IVB, the chip's Uncore frequency was set to match the frequency of the fastest
active core. Because the Uncore contains L3 segments, ring interconnect, and
memory controllers, L3 and memory bandwidth is a function of core frequency on
SNB/IVB. This effect can be observed when setting the frequency to 1.2\,GHz on
IVB (magenta line in Fig~\ref{fig:chip-perf}a): Despite providing the best
energy-to-solution, performance is hurt badly. On HSW, the Uncore is clocked
independently. This leads to a situation in which (due to frequency-induced lower
per-core performance) more cores are required to saturate main memory bandwidth
(cf. magenta line in Fig.\ref{fig:chip-perf}b); however, the lower per-core
power consumption translates into overall energy savings of 11\%, lowering
energy-to-solution to 24.0\,J. These results indicate that energy consumption
could be further reduced if the chips offered frequencies below their current 1.2\,GHz
floor.

Table~\ref{tab:multi-core-results} contains a summary of the final result
pertaining to energy saving for each microarchitecture. The ``reference'' value
corresponds to energy consumption of the baseline implementation running on all
cores clocked at the chip's nominal frequency. Energy-to-solution of the most
energy-efficient version is listed in the ``optimized'' column along with the
number of active cores and their frequency that was used to obtain the results
in the ``configuration'' column.


\begin{figure}[!tb]
    \centering
    \includegraphics[width=0.49\linewidth]{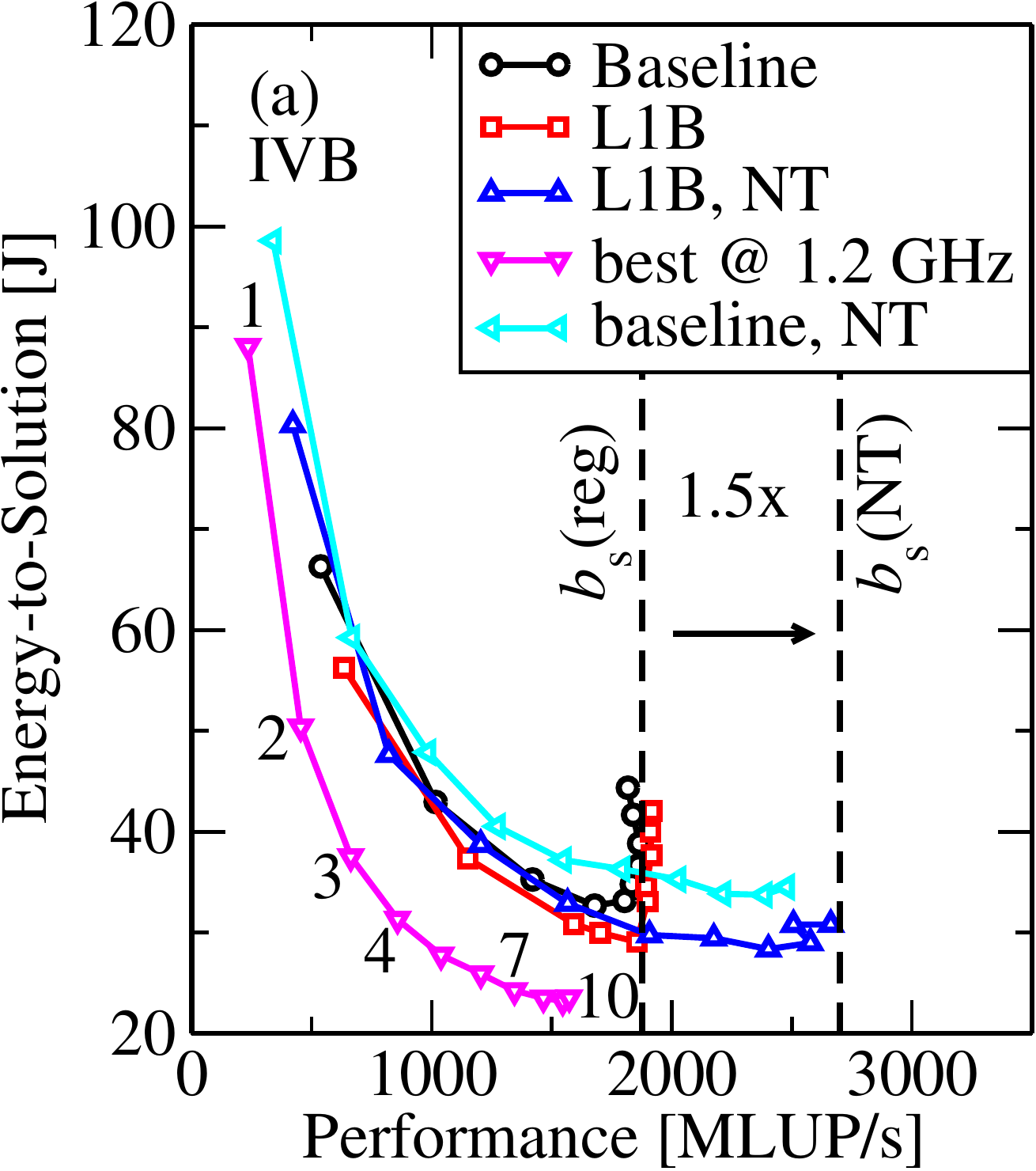}\hfill
    \includegraphics[width=0.49\linewidth]{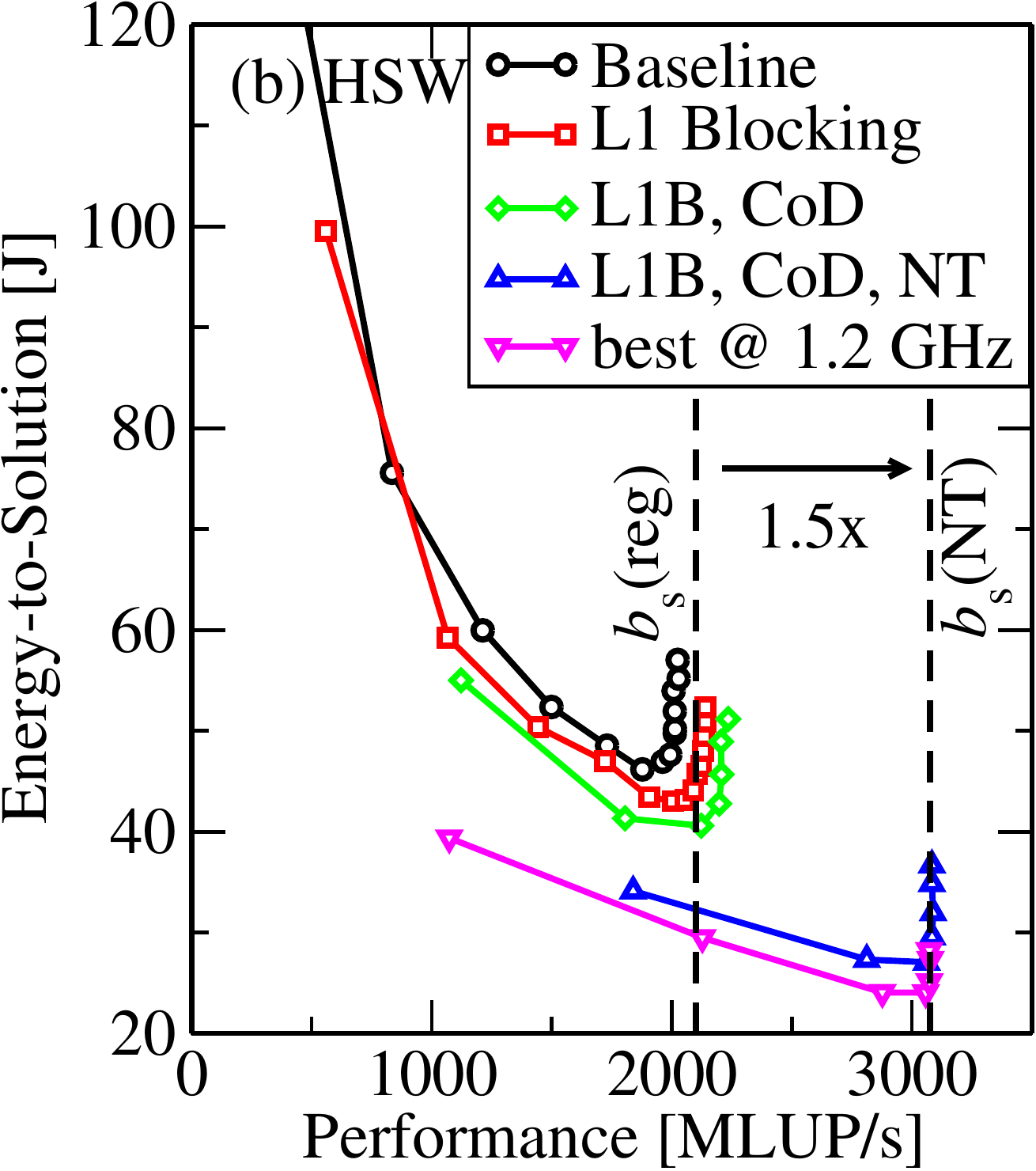}
    \caption{\label{fig:chip-perf}Performance vs. energy-to-solution for different core counts on IVB (a) and HSW (b).}
\end{figure}




\section{Conclusion}
\label{sec:conc}

We have applied a new energy-optimization approach to a 2D Jacobi solver and
analyzed its effects on a range of recent Intel multi-core chips. Using the
execution-cache-memory (ECM) model single-core software improvements were
described and their accuracy validated by measurements. For the first time, the
ECM model has been (a) extended to incorporate non-temporal (NT) stores and (b)
applied to a Broadwell-EP chip. We found energy consumption can reduces by a
factor of 2.1$\times$ on Sandy Bridge-EP, 2.0$\times$ on Ivy Bridge-EP,
2.4$\times$ on Haswell-EP, and 2.3$\times$ on Broadwell-EP. Further, we found
that while NT stores can increase performance on Sandy- and Ivy Bridge-based E5
processors, a direct positive impact on energy consumption could not be
observed; only in combination with frequency tuning do NT stores offer a better
energy-to-solutions on these architectures.  Measurements indicate that this
problem has been solved on Haswell-EP and Broadwell-EP. Moreover, our results
indicate that future microarchitectures that keep core and Uncore frequencies
decoupled could offer improved energy-efficiency if core frequencies below the
current 1.2\,GHz floor were available.  Beyond these immediate results we have
demonstrated the viability of our energy-optimization approach.


\bibliographystyle{abbrv}
\footnotesize{

}

\begin{table}[!tb]
\centering
\caption{\label{tab:multi-core-results}Summary of chip-level benchmarks. Energy improvements shown in parenthesis.}
\resizebox{\linewidth}{!}{%
\begin{tabular}{|c|c|c|c|}
\hline
\textmu{}arch   & Reference     & Optimized             & Configuration \\ \hline
SNB             & 58.7\,J       & 28.0\,J (2.1$\times$) & 7 cores, 1.2\,GHz \\  \hline
IVB             & 44.4\,J       & 22.3\,J (2.0$\times$) & 9 cores, 1.5\,GHz \\  \hline 
HSW             & 57.0\,J       & 24.0\,J (2.4$\times$) & 4 cores, 1.2 GHz \\  \hline
BDW             & 47.9\,J       & 20.4\,J (2.3$\times$) & 4 cores, 1.2\,GHz \\  \hline
\end{tabular}}
\end{table}

\end{document}